\documentclass[submission, Phys]{SciPost}

\usepackage[utf8]{inputenc} 
\usepackage[T1]{fontenc} 	
\usepackage[english]{babel} 

\usepackage[bitstream-charter]{mathdesign}
\usepackage{geometry} 		
\usepackage{amsmath} 		
\usepackage{mathtools} 		
\usepackage{float} 			
\usepackage{graphicx} 		
\usepackage{tabularx} 		
\usepackage{booktabs} 		
\usepackage{color, xcolor} 	
\usepackage{pdfpages} 		
\usepackage{extarrows} 		
\usepackage{multirow} 		
\usepackage{multicol} 		
\usepackage{enumitem} 		
\usepackage{xspace} 		
\usepackage{stackrel} 		
\usepackage{tikz} 			
\usepackage{braket} 		
\usepackage{bm} 			
\usepackage{tensor} 		
\usepackage{slashed} 		
\usepackage{siunitx} 		
\usepackage{lastpage} 		
\usepackage{cite} 			
\usepackage[normalem]{ulem} 
\usepackage{fontawesome} 	
\usepackage{tocloft} 		
\usepackage{titlesec} 		
\usepackage{doi} 			
\usepackage{hyperref} 		
\usepackage[most]{tcolorbox} 					
\usepackage[nameinlink, capitalize]{cleveref} 	
\usepackage[nottoc, notlot, notlof]{tocbibind} 	
\usepackage[ruled, vlined]{algorithm2e} 		
\usepackage{makecell}
\usepackage[makeroom]{cancel}
\usepackage{feynmf}

\makeatletter
\def\BState{\State\hskip-\ALG@thistlm}
\makeatother

\makeatletter
\@ifundefined{pdfoutput}{}{\DeclareGraphicsRule{*}{mps}{*}{}}
\makeatother

\makeatletter
\DeclareRobustCommand*{\bfseries}{%
   \not@math@alphabet\bfseries\mathbf
   \fontseries\bfdefault\selectfont
   \boldmath
}
\makeatother

\hypersetup{
	pdftitle={Economical Jet Taggers},
	pdfauthor={Petitjean et al.},
	colorlinks=true, 			
	linkcolor={red!50!black}, 	
	citecolor={blue!50!black}, 	
	urlcolor={blue!80!black} 	
} 

\DeclareSymbolFont{usualmathcal}{OMS}{cmsy}{m}{n}
\DeclareSymbolFontAlphabet{\mathcal}{usualmathcal}

\SetArgSty{textnormal}
\SetKwComment{Comment}{{\small\#}~}{}
\SetCommentSty{mycommfont}

\setitemize{itemsep=0pt, parsep=0pt} 				
\setenumerate{itemsep=0pt, parsep=0pt} 				
\setitemize{itemsep=2pt,topsep=2pt,parsep=0pt,partopsep=0pt,leftmargin=*}
\setenumerate{itemsep=0pt,topsep=2pt,parsep=0pt,partopsep=0pt,labelindent=3pt,leftmargin=*}
\newlist{todolist}{itemize}{2}
\setlist[todolist]{label=$\square$}
\usepackage{pifont}

\usepackage{amsmath}
 
\usepackage{amsthm} 		
\theoremstyle{definition}

\definecolor{red_cb}{HTML}{e41a1c}
\definecolor{blue_cb}{HTML}{377eb8}
\definecolor{green_cb}{HTML}{4daf4a}
\definecolor{purple_cb}{HTML}{984ea3}
\definecolor{orange_cb}{HTML}{ff7f00}

\definecolor{EmeraldGreen}{HTML}{1ea78d}
\definecolor{EnglishRed}{HTML}{b02427}
\hypersetup{colorlinks=true,urlcolor=EmeraldGreen,citecolor=EmeraldGreen,linkcolor=EnglishRed}

\newcommand{\ie}{\text{i.e.}\;}

\newcommand{\argmin}{\text{argmin}}

\newcommand\one{\leavevmode\hbox{\small1\normalsize\kern-.33em1}}
\newcommand{\sign}{\operatorname{sign}} 	

\newcommand{\loss}{\mathcal{L}} 	

\newcommand{\arXiv}[2][]{%
	\ifthenelse{\equal{#1}{}}%
	{\href{http://arxiv.org/abs/#2}{arXiv:#2}}%
	{\href{http://arxiv.org/abs/#2}{arXiv:#2~[#1]}}}

\newcommand{\gev}{\text{GeV}}

\def\slashchar#1{\setbox0=\hbox{$#1$}           
   \dimen0=\wd0                                 
   \setbox1=\hbox{/} \dimen1=\wd1               
   \ifdim\dimen0>\dimen1                        
      \rlap{\hbox to \dimen0{\hfil/\hfil}}      
      #1                                        
   \else                                        
      \rlap{\hbox to \dimen1{\hfil$#1$\hfil}}   
      /                                         
   \fi}

\newcommand{\tikznode}[2]{%
\ifmmode%
\tikz[remember picture,baseline=(#1.base),inner sep=0pt] \node (#1) {$#2$};%
\else
\tikz[remember picture,baseline=(#1.base),inner sep=0pt] \node (#1) {#2};%
\fi}

\def\mathswitchr#1{\relax\ifmmode{\mathrm{#1}}\else$\mathrm{#1}$\xspace\fi}
\def\mathswitch#1{\relax\ifmmode#1\else$#1$\xspace\fi}

\newcommand{\dz}{\phantom{0}}
\newcommand{\result}[2]{{#1}~\textcolor{error}{$\pm$ {#2}}}

\definecolor{error}{rgb}{0.7, 0.7, 0.7}
\newcommand{\error}[1]{\textcolor{error}{{#1}}}

\graphicspath{{./figs/}}

\usepackage{glossaries}
\usepackage{cleveref}
\usepackage{placeins}

\begin{document}

~\vspace{-0.45in}

\hfill {\footnotesize IPPP/25/93}

\begin{center}{\Large \textbf{
Economical Jet Taggers -- Equivariant, Slim, and Quantized
}}\end{center}

\begin{center}
Antoine Petitjean\textsuperscript{1}, 
Tilman Plehn\textsuperscript{1,2}, 
Jonas Spinner\textsuperscript{3}, and
Ullrich K\"othe\textsuperscript{2}
\end{center}

\begin{center}
{\bf 1} Institut für Theoretische Physik, Universit\"at Heidelberg, Germany \\
{\bf 2} Interdisciplinary Center for Scientific Computing (IWR), Universität Heidelberg, Germany \\
{\bf 3} Institute for Particle Physics Phenomenology, Durham University, UK
\end{center}

\begin{center}
\today
\end{center}


\section*{Abstract}
{\bf 
Modern machine learning is transforming jet tagging at the LHC, but the leading transformer architectures are large, not particularly fast, and training-intensive. We present a slim version of the L-GATr tagger, reduce the number of parameters of jet-tagging transformers, and quantize them. We compare different quantization methods for standard and Lorentz-equivariant transformers and estimate their gains in resource efficiency. We find an order-of-magnitude reduction in energy cost for an moderate performance decrease, down to 1000-parameter taggers. This might be a step towards trigger-level jet tagging with small and quantized versions of the leading equivariant transformer architectures.
}

\vspace{10pt}
\noindent\rule{\textwidth}{1pt}
\tableofcontents\thispagestyle{fancy}
\noindent\rule{\textwidth}{1pt}
\vspace{10pt}

\clearpage
\section{Introduction}
\label{sec:intro}

Modern machine learning (ML) is reshaping the research program at the LHC. Even for standard analyses, triggering, data acquisition, object identification, first-principle simulations, and optimal inference are changing rapidly~\cite{Plehn:2022ftl}. Among the most established ML applications in LHC physics are jet taggers~\cite{Kasieczka:2019dbj,Nachman:2022emq}. Their goal is to utilize the complete information about the jet substructure to determine their partonic nature~\cite{Cogan:2014oua,Baldi:2014kfa,deOliveira:2015xxd,Gallicchio:2010sw,Kasieczka:2017nvn,Qu:2019gqs}. This question leads directly to the input and latent data representations, from image-inspired architectures to graph networks and transformers for permutation-invariant point clouds. In addition, jet taggers were shown to benefit from the fundamental Lorentz symmetry~\cite{Butter:2017cot}. This motivated a series of Lorentz-equivariant graph-networks taggers~\cite{Gong:2022lye,Qiu:2022xvr,Bogatskiy:2022czk,Qiu:2023ihi,ruhe2023clifford}, where in physics we would use the term `covariant' rather than `equivariant'. The final step of the development of equivariant taggers is marked by Lorentz-equivariant transformers with learned symmetry breaking, using either a geometric algebra representation~\cite{Brehmer:2024yqw} or symmetry-linked local reference frames for each constituent~\cite{Favaro:2025pgz}. Their advantage over the corresponding graph networks is that transformers benefit significantly more from larger training datasets and scale more efficiently with network size.

An alternative path to high-performance taggers is pre-training of larger and larger transformers, with the implicit assumption that they learn the underlying physics structures~\cite{Qu:2022mxj,Brehmer:2024yqw,Favaro:2025pgz,Bhimji:2025isp,Birk:2025fbs}. Upscaling networks and training datasets is inspired by industry applications and has become the standard method to roll out increasingly powerful ML-tools. However, in LHC physics we have to account for many limiting factors, which can include the memory usage of the standard analysis hardware or specific hardware used for event triggering. For triggering, jet classification does not yet play a role, but it definitely should~\cite{Loncar:2020hqp,Hawks:2021ruw,Krause:2025qnl,Rai:2025cog}. This leads us to asking the inverse question: \textsl{How small and how efficient can we make modern jet taggers with limited performance loss?}.

The first and immediate question is how small and how fast we can make an equivariant transformer, assuming that symmetry-aware data representations will improve training and performance under constraints. In Sec.~\ref{sec:slim}, we present and benchmark a streamlined version of the L-GATr architecture that uses a more efficient latent representation based only on scalars and vectors. This slim Lorentz-equivariant tagger can be applied to many LHC tasks, from jet tagging to amplitude regression and event generation. In Sec.~\ref{sec:slim-mini} we compare the performance of L-GATr-slim and LLoCa taggers~\cite{Favaro:2025pgz} down to 1000-parameter networks, either with a reduced number of layers or with a reduced numbers of parameters per layer.

The comparison between the L-GATr-slim and LLoCa taggers provides us with a baseline for further computational optimization in Sec.~\ref{sec:quantization}. In this second step we quantize the equivariant taggers. Optimized data types are an established strategy for large language models, using different precision for different network layers to minimize computational cost and energy consumption. In Sec.~\ref{sec:weight-quantization} we introduce weight quantization of the linear layers via a regularization constraint implemented with the proximal gradient method, leading to piecewise-affine regularized quantization (PARQ), and straight-through estimation (STE) as a special case. We apply this quantization to Lorentz-equivariant taggers in Sec.~\ref{sec:gpu-tagging}. We find that they can be quantized like standard networks, leading to a significant efficiency gain with minimal performance loss. Finally, in Sec.~\ref{sec:fpga-tagging} we combine both strategies to develop truly minimal, \ie small and quantized versions of the leading tagger architectures. Such maximally resource-efficient versions might become useful for triggering at the HL-LHC.

\section{L-GATr-slim}
\label{sec:slim}

The standard data representation for neural networks applied to LHC physics is point clouds, including 4-vectors for each particle, jet constituent, or partons. The two symmetries governing these 4-vectors are permutations and Lorentz transformations. Graph-based and transformer architectures ensure permutation invariance. Explicit Lorentz equivariance with learned symmetry breaking in the network architecture has been shown to lead to significant performance improvements for graph networks~\cite{Gong:2022lye,Qiu:2022xvr,Bogatskiy:2022czk,Qiu:2023ihi,ruhe2023clifford} and for transformers~\cite{Brehmer:2024yqw,Favaro:2025pgz}. The key advantage of transformers over other graph network architectures is their superior scaling: they train efficiently on large datasets thanks to highly parallelizable attention, often yielding the best performance as data and network size grow.

There exist two fundamentally different implementations of Lorentz-equivariant transformers. First, L-GATr~\cite{Brehmer:2024yqw} expresses input and output using a geometric algebra representation and replaces the transformer layers with operations that keep track of scalar, pseudo-scalar, vector, axial-vector, and antisymmetric rank-two tensor components. Second, LLoCa~\cite{Favaro:2025pgz} assigns a local reference frame to each 4-vector, represents latent features as invariants in these frames, and incorporates Lorentz transformations into the message passing. Both architectures  give comparable performance on standard LHC tasks, such as amplitude regression, top tagging, and event generation. As the pseudo-scalar, axial-vectors, and tensor representations in L-GATr are not needed for most LHC applications~\cite{Qiu:2023ihi}, we develop a slim L-GATr architecture that only uses scalars and vectors.

\subsection{Lorentz-scalars and vectors}
\label{sec:slim-architecture}

The L-GATr-slim architecture generalizes standard transformer building blocks to operate on coupled scalar $s$ and vector $v$ representations:
\begin{itemize}
\item 
First, we extend the scalar linear layer by a linear operation acting on vectors, where all vector components are multiplied by the same learnable scalar coefficient. Different scalar coefficients for the vector components would violate Lorentz equivariance. This operation is equivalent to the L-GATr linear layer,
\begin{align}
    &\text{Linear} (x) 
    = 
    \text{Linear} 
    \begin{pmatrix} s\\v \end{pmatrix} 
    = \begin{pmatrix}\text{Linear}_s(s)\\\text{Linear}_v(v)\end{pmatrix}
    = \begin{pmatrix} w_s\cdot s + b_s \\ w_v\cdot v \end{pmatrix} \\
    &\text{with} \qquad 
    s\in\mathbb{R}^{c_s^\text{in}} \quad
    w_s\in\mathbb{R}^{c_s^\text{out}\times c_s^\text{in}} \quad
    b_s\in\mathbb{R}^{c_s^\text{out}} \quad
    v\in\mathbb{R}^{c_v^\text{in}\times 4} \quad
    w_v\in \mathbb{R}^{c_v^\text{out}\times c_v^\text{in}} \; .\notag 
\end{align}

\item 
For the nonlinearity we extend the Gated Linear Unit (GLU)~\cite{dauphin2017language}. Instead of directly applying the GELU nonlinearity to the linear layer output, GLU multiplies these activations with the output of a second linear layer. We adapt this idea to the vector channel by applying the nonlinearity to the inner product of two vectors, and multiply it by the vector output of the L-GATr-slim linear layer,
\begin{align}
    \text{GLU}\begin{pmatrix}s\\v\end{pmatrix} = \begin{pmatrix}
        \text{GELU}\left(\text{Linear}_{s,1}(s)\right) \; \text{Linear}_{s,2} (s) \\
        \text{GELU}\left(\langle \text{Linear}_{v,1}(v),\text{Linear}_{v,2}(v)\rangle\right) \; \text{Linear}_{v,3}(v)
    \end{pmatrix} \; .
\end{align}
$\text{Linear}_{s,j}$ and $\text{Linear}_{v,j}$ denote different linear layers, but implemented as one fused linear operation with an increased output dimensionality, by a factor two for scalars and a factor three for vectors. When using GLU we follow the common practice of increasing the number of output channels by another factor of two compared to the number of input channels.

\item 
Next, we modify the standard RMSNorm to account for the geometric properties of vector channels. We use the absolute value of the Minkowski inner product to evaluate the vector contribution to the norm. The absolute value is required, because the Minkowski inner product can become negative. This is very similar to L-GATr, with the only difference that L-GATr normalizes multivectors and scalars seperately,
\begin{align}
    \text{RMSNorm}\begin{pmatrix}s\\v\end{pmatrix} = \left(\frac{1}{c_v}\sum_{c=1}^{c_v} \left| \langle v_c,v_c\rangle \right|^2 + \frac{1}{c_v}\sum_{c=1}^{c_s} s_c^2 +\epsilon\right)^{-1/2} \begin{pmatrix}s \\ v\end{pmatrix} \; .
\end{align}

\item 
Finally, we construct the scalar attention matrix closely following L-GATr, 
\begin{align}
    \text{Attention}\begin{pmatrix}q_s,k_s,v_s\\q_v,k_v,v_v\end{pmatrix}_i = \sum_{j=1}^{n_t} \text{Softmax}_j\left(\frac{\sum_{c=1}^{c_s} q_{s,ic}k_{s,jc} + \sum_{c=1}^{c_v} \langle q_{v,ic},k_{v,jc}\rangle}{\sqrt{4n_v + n_s}}\right) \begin{pmatrix}v_{s}\\v_{v}\end{pmatrix}_j \; .
\end{align}
The inner product is implemented as a list of prefactors multiplied onto the query vector before evaluating the (euclidean) inner product using optimized attention kernels. The token dimensionality is $n_t$.
\end{itemize}
We combine these layers into the L-GATr-slim transformer architecture,
\begin{align}
    \bar x &= \text{RMSNorm}(x) \notag\\
    \text{AttentionBlock}(x) &= \text{Linear}\circ\text{Attention}\left(\text{Linear}(\bar x),\text{Linear}(\bar x), \text{Linear}(\bar x)\right)+x\notag\\
    \text{MLPBlock}(x) &= \text{Linear}\circ \text{GLU}(\bar x) + x\notag\\
    \text{Block}(x) &= \text{MLPBlock}\circ\text{AttentionBlock}(x)\notag\\
    \text{L-GATr-slim}(x) &= \text{Linear}\circ\text{Block}\circ\text{Block}\circ\cdots\circ\text{Block}\circ\text{Linear}(x)
\end{align}
The main difference to L-GATr is the reduced number of hidden components/channels and the fact that our L-GATr-slim implementation can be compiled with \texttt{torch.compile} to increase computing efficiency. Moreover, L-GATr-slim uses slightly different designs for the normalization and nonlinearity that simplify the processing of scalars and vectors through the network without sacrificing performance. We find that it matches the L-GATr performance even without an outer product, so we drop this operation to save significant extra computational cost.

\subsection{Jet tagging}
\label{sec:slim-tagging}

Because our focus will be jet tagging, we first show the L-GATr-slim performance for the standard top tagging benchmark~\cite{Kasieczka:2019dbj}. We compare our slim architecture to other leading architectures in Tab.~\ref{tab:top_tagging} and find     that it matches the performance of L-GATr, the LLoCa-Transformer, and different Lorentz-equivariant graph networks. The hyperparameters for the full-sized L-GATr-slim are given in App.~\ref{app:hyperparams}. For details on the established networks and their hyperparameters we refer to Ref.~\cite{Favaro:2025pgz}. Because detector effects violate Lorentz symmetry, the tagger has to be able to incorporate the corresponding symmetry breaking. To allow for that, we keep the beam and time axes as additional input particles~\cite{Brehmer:2024yqw,Favaro:2025pgz}.

\begin{figure}[t]
    \includegraphics[width=0.495\linewidth]{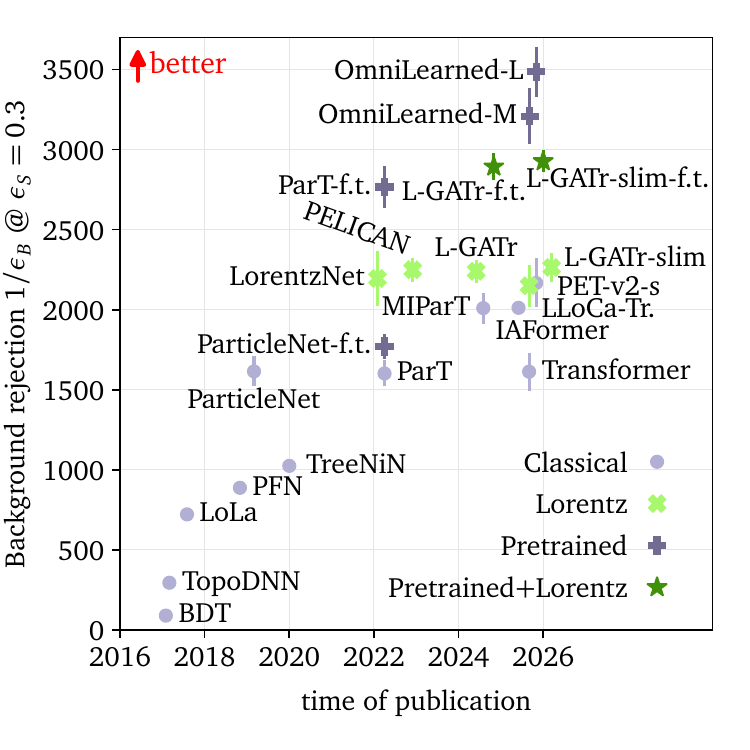}
    \includegraphics[width=0.495\linewidth,page=1]{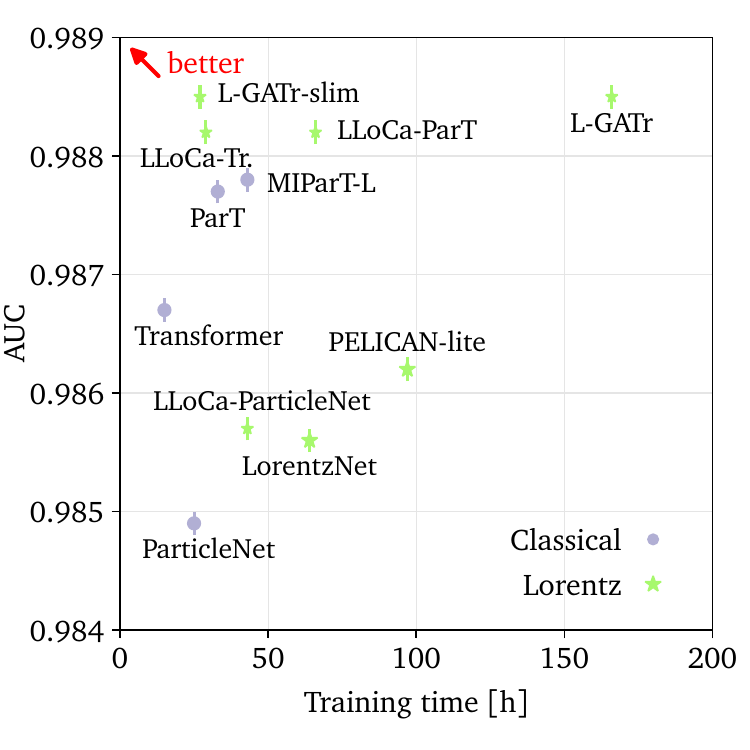}
    \caption{Left: Progress in top tagging through advanced network architectures over time. Right: Efficiency of tagging architectures on the JetClass dataset.}
    \label{fig:tagger-money}
\end{figure}

\begin{table}[b!]
    \centering
    \footnotesize
    \begin{tabular}{l c lllllr}
        \toprule
        Network && Accuracy & AUC & $1/\epsilon_B$ ($\epsilon_S=0.5$) & $1/\epsilon_B$ ($\epsilon_S=0.3$) & \# Params & \# Train\\
        \midrule
        ParticleNet \cite{Qu:2019gqs} && 0.940\dz{} & 0.9858 & \result{397}{\dz{}7} & \result{1615}{\dz{}93} & 0.4M & 1.2M \\
        Transformer \cite{Favaro:2025pgz} && 0.9393 & \result{0.9855}{0.0001} & \result{389}{\dz{}6} & \result{1613}{118} & 2.0M & 1.2M \\
        ParT \cite{Qu:2022mxj} && 0.940\dz{} & 0.9858 & \result{413}{16} & \result{1602}{\dz{}81} & 2.1M & 1.2M \\
        ParT (longer) && 0.9416\dz{} & 0.9865 & \result{485}{15} & \result{1808}{\dz{}33} & 2.1M & 1.2M \\
        MIParT \cite{Wu:2024thh} && 0.942\dz{} & 0.9868 & \result{505}{\dz{}8} & \result{2010}{\dz{}97} & 2.2M & 1.2M \\
        IAFormer \cite{Esmail:2025kii} && 0.942 & 0.987 & \result{510}{\dz{}6} & \result{2012}{\dz{}30} & 0.2M & 1.2M \\
        PET v2-s \cite{Bhimji:2025isp} && 0.943 & 0.987 & \result{505}{14} & \result{2167}{153} & 3.0M & 1.2M \\
        LorentzNet* \cite{Gong:2022lye} && 0.942\dz{} & 0.9868 & \result{498}{18} & \result{2195}{173} & 0.2M & 1.2M \\
        PELICAN* \cite{Bogatskiy:2023nnw} && 0.9426 & \result{0.9870}{0.0001} & -- & \result{2250}{\dz{}75} & 0.2M & 1.2M \\
        CGENN* \cite{ruhe2023clifford} &&  0.942\dz{} & 0.9869 & 500 & 2172 & 0.3M & 1.2M \\
        LLoCa-Transformer* \cite{Favaro:2025pgz} && 0.9416 & \result{0.9866}{0.0001} & \result{492}{15} & \result{2150}{130} & 2.0M & 1.2M \\
        L-GATr* \cite{Brehmer:2024yqw} && 0.9423 & \result{0.9870}{0.0001} & \result{540}{20} & \result{2240}{\dz{}70} & 1.1M & 1.2M \\
        L-GATr-slim* && 0.9420\dz{} & \result{0.9869}{0.0001} & \result{546}{\dz{7}} & \result{2264}{\dz{}93} & 2.0M & 1.2M \\
        \midrule
        ParticleNet-f.t. \cite{Qu:2022mxj} && 0.942\dz{} & 0.9866 & \result{487}{\dz{}9} & \result{1771}{\dz{}80} & 0.4M & 100M \\
        OmniLearn \cite{Mikuni:2024qsr} && 0.942 & 0.9872 & \result{568}{\dz{}9} & \result{2647}{192} & 2.0M & 100M \\
        ParT-f.t. \cite{Qu:2022mxj} && 0.944\dz{} & 0.9877 & \result{691}{15} & \result{2766}{130} & 2.1M & 100M \\
        MIParT-f.t. \cite{Wu:2024thh} && 0.944\dz{} & 0.9878 & \result{640}{10} & \result{2789}{133} & 2.3M & 100M \\
        L-GATr-f.t.* \cite{Brehmer:2024yqw} && 0.9446 & 0.9879\dz{} & \result{651}{11} & \result{2894}{\dz{}84} & 1.1M & 100M \\
        L-GATr-slim-f.t.* && 0.9442 & 0.9879 & \result{655}{\dz{}5} & \result{2927}{\dz{}70} & 2.0M & 100M \\
        OmniLearned-M \cite{Bhimji:2025isp} && 0.944 & 0.9880 & \result{656}{12} & \result{3208}{176} & 58M & 1058M \\
        OmniLearned-L \cite{Bhimji:2025isp} && 0.944 & 0.9880 & \result{688}{\dz{}9} & \result{3486}{157} & 423M & 1058M \\
        \bottomrule
    \end{tabular}
    \caption{Top tagging accuracy, AUC and background rejection rates for two fixed signal efficiencies~\cite{Kasieczka:2019dbj}. Apart from L-GATr-slim and L-GATr-slim-f.t., we added 'ParT (longer)', a ParT training using the same training hyperparameters as L-GATr-slim. We indicate Lorentz-equivariance with an asterisk and estimate uncertainties using five trainings. The first set is  trained from scratch, the second using pretraining.}
    \label{tab:top_tagging}
\end{table}

\begin{table}[t]
\centering
\begin{small}
\begin{tabular}{lll S[table-format=3.0] @{\hspace{0.4cm}} S[table-format=4.0] @{\hspace{0.4cm}} S[table-format=2.1] @{\hspace{0.4cm}} S[table-format=3.0]}
\toprule
Network & Accuracy & AUC & \text{Time} & \text{FLOPs} & \text{Memory} & \text{Parameters} \\
\midrule
MIParT-L \cite{Wu:2024thh}      & 0.861\dz{} & 0.9878\dz{} & 43h & 225M & 53.6G & 2380k \\
LorentzNet* \cite{Gong:2022lye} & 0.847\dz{} & 0.9856\dz{} & 64h & 676M & 20.5G & 223k \\
PELICAN-lite*~\cite{Favaro:2025pgz} & 0.851\dz{}   & 0.9862\dz{} & 97h & 1370M & 27.4G & 244k \\
ParticleNet \cite{Qu:2019gqs}            & 0.844\dz{} & 0.9849\dz{} & 25h & 413M & 16.5G & 366k \\
LLoCa-ParticleNet* \cite{Favaro:2025pgz} & 0.848 & 0.9857 & 43h & 517M & 23.5G & 385k \\
ParT \cite{Qu:2022mxj} & 0.861\dz{} & 0.9877\dz{} & 33h & 211M & 13.3G & 2141k \\
LLoCa-ParT* \cite{Favaro:2025pgz} & 0.864 & 0.9882 & 66h & 315M & 19.9G & 2160k \\
Transformer \cite{Favaro:2025pgz} & 0.855 & 0.9867 & 15h & 210M & 2.3G & 1979k \\
LLoCa-Transformer* \cite{Favaro:2025pgz} & 0.864 & 0.9882  & 28h & 219M & 4.1G & 1980k \\
L-GATr* \cite{Brehmer:2024yqw} & 0.866\dz{} & 0.9885\dz{} & 166h & 2060M & 19.0G & 1079k \\
L-GATr-slim* & 0.866 & 0.9885 & 27h & 329M & 8.1G & 2031k \\
\bottomrule
\end{tabular}
\end{small}
\caption{Performance and computational cost for multi-class taggers on the JetClass dataset~\cite{Qu:2022mxj}. We show accuracy, AUC averaged over all pairs of classes, time for a complete training on a H100 GPU, FLOPs per forward pass, maximum memory consumption during training, and number of learnable parameters. See Table~\ref{tab:tagging_rejection} for the corresponding rejection rates. Lorentz-equivariant networks are denoted with an asterisk. We use a narrower Frames-Net for the LLoCa-Transformer, because it achieves equal performance at reduced cost.}
\label{tab:jetclass_tagging}
\end{table}

\begin{table}[t]
    \fontsize{9}{9}\selectfont
    \centering
    \begin{tabular}{lccccccccccc}
        \toprule
        & $H \to b\bar{b}$ & $H \to c\bar{c}$ & $H \to gg$ & $H \to 4q$ & $H \to l\nu q\bar{q}'$ & $t \to b q\bar{q}'$ & $t \to b l\nu$ & $W \to q\bar{q}'$ & $Z \to q\bar{q}$ \\        
        & Rej$_{50\%}$ & Rej$_{50\%}$ & Rej$_{50\%}$ & Rej$_{50\%}$ & Rej$_{99\%}$ & Rej$_{50\%}$ & Rej$_{99.5\%}$ & Rej$_{50\%}$ & Rej$_{50\%}$ \\
        \midrule
        MIParT-L~\cite{Wu:2024thh} & 10753 & 4202 & 123 & 1927 & 5450 & 31250 & 16807 & 542 & 402 \\
        LorentzNet*~\cite{Favaro:2025pgz} & 8475 & 2729 & 111 & 1152 & 3515 & 13889 & 10257 & 400 & 303\\
        PELICAN-lite*~\cite{Favaro:2025pgz} & 8333 & 3040 & 113 & 1321 & 3802 & 17084 & 10363 & 435 & 332\\
        ParticleNet~\cite{Qu:2019gqs} & 7634 & 2475 & 104 & 954 & 3339 & 10526 & 11173 & 347 & 283 \\
        LLoCa-ParticleNet*~\cite{Favaro:2025pgz} & 7463 & 2833 & 105 & 1072 & 3155 & 10753 & 9302 & 403 & 306 \\
        ParT~\cite{Qu:2022mxj} & 10638 & 4149 & 123 & 1864 & 5479 & 32787 & 15873 & 543 & 402 \\
        LLoCa-ParT*~\cite{Favaro:2025pgz} & 11561 & 4640 & 125 & 2037 & 5900 & 41667 & 19231 & 552 & 419 \\
        Transformer~\cite{Favaro:2025pgz} & 10753 & 3333 & 116 & 1369 & 4630 & 24390 & 17857 & 415 & 334 \\
        LLoCa-Transformer*~\cite{Favaro:2025pgz} & 11628 & 4651 & 125 & 2037 & 5618 & 39216 & 17241 & 548 & 410 \\
        L-GATr*~\cite{Brehmer:2024yqw} & 12987 & 4819 & 128 & 2311 & 6116 & 47619 & 20408 & 588 & 432 \\
        L-GATr-slim* & 12121 & 4843 & 127 & 2309 & 5952 & 40113 & 21053 & 585 & 429 \\
        \bottomrule
    \end{tabular}
    \caption{Background rejection rates $1/\epsilon_B$ for the JetClass dataset~\cite{Qu:2022mxj}. See Table~\ref{tab:jetclass_tagging} for additional metrics. Lorentz-equivariant networks are denoted with an asterisk. }
    \label{tab:tagging_rejection}
\end{table}

Next, we benchmark L-GATr-slim on the larger JetClass dataset, covering light quark or gluon jets, $Z$, $W$, and $H$ jets, as well as top jets. The dataset contains 10M training jets for each of the ten different jet labels. We report the AUC averaged over all pairs of classes, accuracy, rejection rates as well as computational cost metrics in Tables~\ref{tab:jetclass_tagging} and~\ref{tab:tagging_rejection}. Again, L-GATr-slim reproduces the performance of other leading Lorentz-equivariant transformers. In addition, L-GATr-slim has the same computational cost as the LLoCa-Transformer and is significantly more efficient than L-GATr, in terms of FLOPs, memory consumption, and training time.

The development of modern jet taggers from the classic 2017 BDT is illustrated in the left panel of Fig.~\ref{fig:tagger-money}. We note that the highest-performing taggers, denoted as f.t. for fine-tuned, are pretrained on increasingly large datasets, as can be seen in Tab.~\ref{tab:top_tagging}. In the right panel of Fig.~\ref{fig:tagger-money} we show the same taggers, measuring the performance in AUC for a given training time on a single H100 GPU. This shows that performance is not automatically correlated with computing cost and motivates the development of resource-efficient networks for LHC analyses.

\subsection{Amplitude regression}
\label{sec:slim-amplitudes}

As we view Lorentz-equivariant transformers as tools to construct physics-motivated and efficient latent representations for a broad range of LHC applications, we move to the fully supervised regression of partonic transition amplitudes as a second benchmark task. Partonic transition amplitudes can be expressed analytically and exactly as functions of 4-momenta, but they have to be Lorentz-invariant. Reducing the dimensionality of the input significantly increases the accuracy of the surrogate network, which becomes a key advantage for expensive loop amplitudes and/or amplitudes with many external particles~\cite{Aylett-Bullock:2021hmo,Maitre:2021uaa,Badger:2022hwf,Maitre:2023dqz,Brehmer:2024yqw,Breso:2024jlt,Bahl:2024gyt,Bahl:2025xvx,Villadamigo:2025our,Beccatini:2025tpk}. 

Our benchmark process for amplitude regression is 
\begin{align}
    q\bar q \to Z + n g \qquad \text{with} \qquad  n = 1... 4 \; ,
\end{align}
with the acceptance cuts
\begin{align}
    p_T > 20\;\text{GeV}\qquad \text{and}\qquad \Delta R>0.4 \;,
\end{align}
as described in detail in Ref.~\cite{Favaro:2025pgz}. The size of the training dataset is 10M amplitudes. The MSE loss evaluates standardized logarithmic amplitudes over phase space points $x$,
\begin{align}
    \loss = \left| \mathcal{A}_\text{NN}(x) - \mathcal{A}_\text{true}(x) \right|^2
    \qquad \text{with} \qquad 
    \mathcal{A}(x) = \frac{\log A(x) - \overline{\log A}}{\sigma_{\log A}} \; .
\end{align} 
For simplicity, we use the same MSE as the performance metric for the $Z + 4$-gluon case in Tab.~\ref{tab:amp_multiplicity}. In the corresponding Figure we show the same outcome in terms of the relative accuracy
\begin{align}
    \Delta(x) = \frac{A_\text{NN}(x) - A_\text{true}(x) }{A_\text{true}(x) } \; ,
\end{align}
histogrammed over an independent test dataset. We compare the Lorentz-equivariant transformer architectures to an MLP with 4-vectors and invariants as input (MLP-I), a standard and a Lorentz-equivariant graph network (GNN and LLoCa-GNN), and a vanilla transformer with 4-vector and PID inputs. As for the other tasks, L-GATr-slim matches the performance of the full L-GATr and the LLoCa-Transformer architectures, outperforming the different graph networks and the vanilla transformer. However, L-GATr-slim requires 20 times fewer training operations and takes less than half of the training time compared to the full L-GATr. With this, L-GATr-slim is almost on par with the resource-efficient and yet accurate LLoCa-Transformer.

\begin{table}[t]
    \begin{minipage}{0.52\linewidth}
    \begin{small}
    \begin{tabular}{l S[table-format=3.1] @{\ \error{$\pm$} } lrr}
    \toprule
     Network  & \multicolumn{2}{c}{MSE$\times 10^{-5}$} & FLOPs & \text{Time} \\ \midrule
     MLP-I~\cite{Brehmer:2024yqw} & 137.0 & \error{2} & 0.1M & 0.4h \\
     GNN~\cite{Favaro:2025pgz}& 10.5 & \error{0.2} & 20.7M & 0.9h \\
     LLoCa-GNN \cite{Favaro:2025pgz}& \phantom{00}5.0 & \error{0.2} & 22.3M & 1.5h \\
     Transformer \cite{Favaro:2025pgz}& 8.3 & \error{0.3} & 14.9M & 1.3h \\
     LLoCa-Transf. \cite{Favaro:2025pgz}& \phantom{00} 1.2 & \error{0.2} & 16.3M & 2.3h \\
     L-GATr~\cite{Brehmer:2024yqw} & 1.8 & \error{0.2} & 528.0M & 8.3h \\
     L-GATr-slim & 1.8 & \error{0.1} & 23.6M & 3.6h \\
     \bottomrule
    \end{tabular}
    \end{small}
    \end{minipage}
    \begin{minipage}{0.48\linewidth}
    \includegraphics[width=\linewidth,page=2]{figs/amplitudes.pdf}
    \end{minipage}
    \caption{
    Performance for $Z+4g$ amplitude regression. Left: MSE performance, cost in terms of FLOPs, and training time on a H100 GPU. Right: relative accuracy over a test dataset. We show error bands based on the standard deviation of three different random seeds.}
    \label{tab:amp_multiplicity}
\end{table}

\subsection{Event generation}
\label{sec:slim-generation}

A third standard task for equivariant transformer representation is the generation of parton-level or jet-level events. Here, a generative network with an equivariant transformer backbone is trained on a set of events from a standard event generator~\cite{Butter:2019cae,Butter:2021csz,Butter:2023fov,Butter:2024zbd}. Ideally, the network can then sample more events than it was trained on, and extremely fast~\cite{Butter:2020qhk,Bahl:2025ryd}. The same kind of event-level generative network is required for neural importance sampling~\cite{Gao:2020vdv,Bothmann:2020ywa,Heimel:2022wyj,Heimel:2023ngj} and generative unfolding~\cite{Bellagente:2019uyp,Bellagente:2020piv,Diefenbacher:2023wec,Huetsch:2024quz,Butter:2025via,Petitjean:2025tgk}.

\begin{figure}[t]
    \includegraphics[width=0.495\linewidth,page=1]{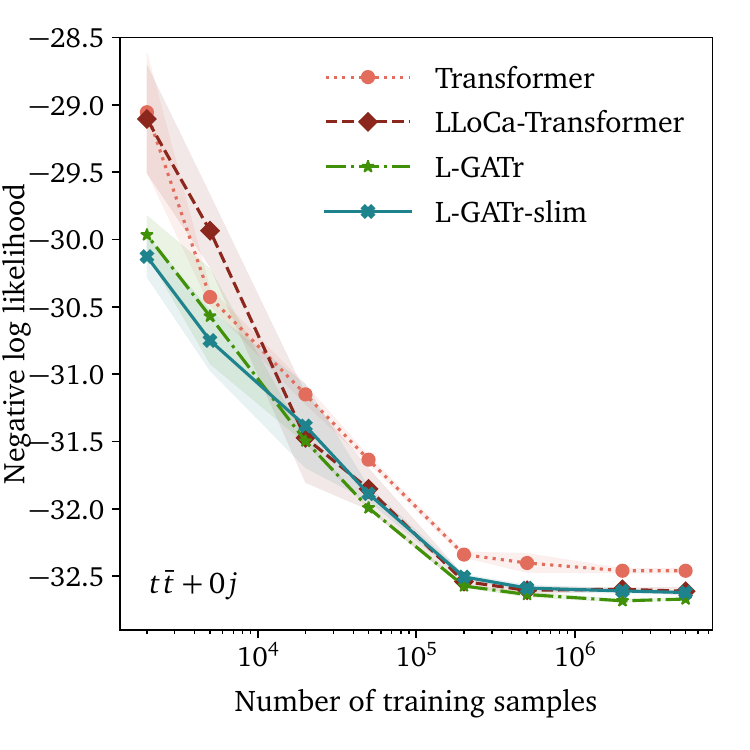}
    \includegraphics[width=0.495\linewidth,page=2]{figs/generation_NLL.pdf}
    \caption{Event generation performance of L-GATr-slim for top pair production with additional jets. We compare it to the original results for L-GATr, Transformer, and LLoCa-Transformer~\cite{Brehmer:2024yqw,Favaro:2025pgz}.}
    \label{fig:eventgen}
\end{figure}

Our benchmark process is two hadronically decaying top quarks in association with up to four extra jets,
\begin{align}
    pp\to t_h \bar t_h + nj,\qquad n=0...4\;,
\end{align}
described in detail in Ref.~\cite{Brehmer:2024yqw}. We require exactly two tagged $b$-jets and apply the acceptance cuts
\begin{align}
 p_{T,j} > 22~\gev 
 \qquad 
 \Delta R_{jj} > 0.5 
 \qquad |\eta_j| < 5 \; .
\end{align} 

Our generative network uses Riemannian flow matching to take into account the phase space topology, and reference vectors to induce symmetry breaking~\cite{Brehmer:2024yqw}. In the left panel of Fig.~\ref{fig:eventgen} we show the negative log-likelihood for all generated events as a function of the training dataset size. As all other shown transformers, L-GATr-slim requires around 500.000 events for optimal performance. The typical percent-level performance in terms of binned standard observables matches the results from Refs.~\cite{Brehmer:2024yqw,Favaro:2025pgz}. The right panel of Fig.~\ref{fig:eventgen} shows the scaling of the generative network accuracy with the phase space dimension. Again, the performance of L-GATr-slim matches the full L-GATr und the LLoCa-Transformer, with no visible degradation.

\subsection{Ultra-mini taggers}
\label{sec:slim-mini}

\begin{figure}[b!]
    \includegraphics[width=0.495\linewidth,page=1]{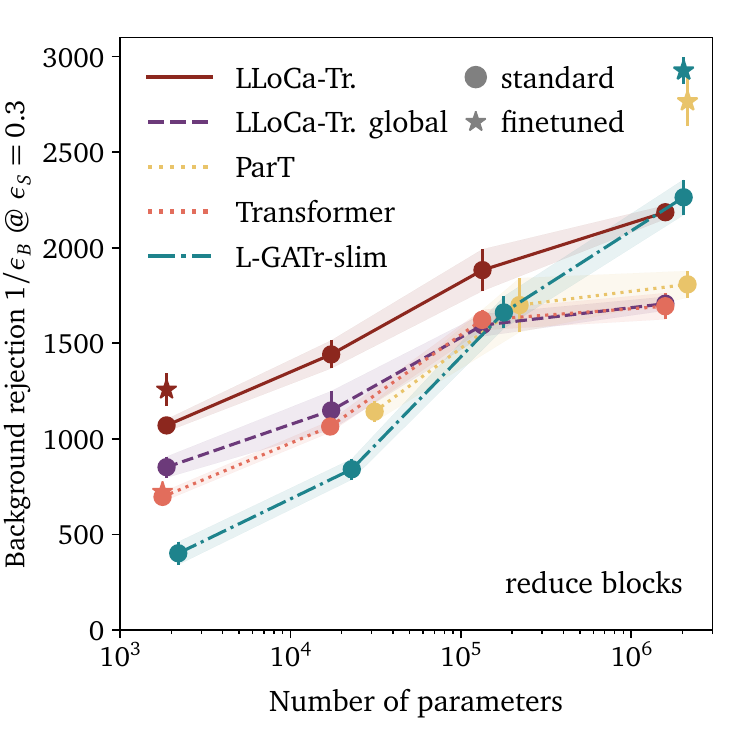}
    \includegraphics[width=0.495\linewidth,page=2]{figs/param_scaling.pdf}
    \caption{Top taggers with decreasing number of parameter, down to 1000. We decrease the number of blocks from 10 to 4, 2, and 1 (left) or keep them fixed at 10 (right). The left panel includes results for taggers pretrained on the JetClass datasets. ParT requires at least two blocks in the standard implementation, so we do not show a 1000-parameter version.}
    \label{fig:scaling}
\end{figure}

Finally, we study equivariant transformers for much fewer parameters. This downscaling is motivated by the potential use of transformers at different trigger levels, where resources are scarce~\cite{Odagiu:2024bkp}. This is why we go back to jet tagging to ask what fraction of network parameters we can spare at limited computational cost. We train a standard transformer, ParT, the LLoCa-Transformer, and L-GATr-slim with a decreasing number of 2M, 200k, 20k, and 2k parameters, see App.~\ref{app:hyperparams}. For even smaller parameter numbers we refer to Refs.~\cite{Bogatskiy:2023fug} and~\cite{Vent:2025ddm}.

In the left panel of Fig.~\ref{fig:scaling} we see that when the number of network blocks and width are decreased jointly, the LLoCa-Transformer emerges as the leading architecture at small network sizes. L-GATr-slim falls behind the other networks, because of the limited number of operations that exchange information between the scalar and vector channels. Pretraining on JetClass and fine-tuning on the top-tagging dataset yields a clear performance gain for the 1000-parameter LLoCa-Transformer due to the increased expressivity, while the 1000-parameter standard transformer sees only a marginal improvement.

The picture changes in the right panel of Fig.~\ref{fig:scaling}, where we keep the number of blocks fixed at 10 and instead reduce the number of parameters per layer. This allows the L-GATr-slim architecture to maintain a background rejection rate above 1000 using only two vector and four scalar channels, marking the leading performance of the 1000-parameter transformer architectures. This shows how we can balance the size and the performance of modern jet taggers, and how the underlying architecture can make a difference in this non-trivial optimization task. For both examples, minimal number of parameters and minimal number of layers, one of the two Lorentz-equivariant architectures works best.

\section{Quantized jet taggers}
\label{sec:quantization}

The performance of reduced-size taggers motivates further steps to reduce computational cost, specifically for the LLoCa-Transformer tagger. The bottleneck operation in neural networks is typically the linear layer, which involves large matrix multiplications. So-called quantization performs these matrix multiplications at reduced precision, such that they become faster and require less memory. For other network operations reduced precision would hurt performance, so they should be kept at bfloat16 or even float32 precision. 

Quantized jet taggers have been studied previously, for example in P-DAT-Bit~\cite{Krause:2025qnl} and BitParT~\cite{Rai:2025cog}. Here we describe how to quantize four representative transformer-based tagging architectures:
\begin{itemize}
    \item Transformer: vanilla transformer;
    \item ParT: Transformer using edge features in attention, and class attention to aggregate over particles;
    \item L-GATr-slim: Lorentz-equivariant transformer, using vector and scalar representations;
    \item LLoCa: Lorentz-equivariant transformer, using Lorentz local canonicalization. We also test a simpler global Lorentz canonicalization.
\end{itemize}

\subsection{Low-precision data types}
\label{sec:input-quantization}

\begin{table}[b!]
    \centering
    \begin{small} 
    \begin{minipage}[t]{0.6\linewidth}
    \centering
    \begin{tabular}{lccc}
        \toprule
        & \multicolumn{3}{c}{teraFLOPS} \\
        & V100~\cite{V100-reference} & A100~\cite{A100-reference} & H100~\cite{H100-reference} \\
        & 2017 & 2020 & 2022 \\
        \midrule
        float64 & 7 & 19.5 & 51 \\
        float32 & 112 & 156 & 756 \\
        float16/bfloat16 & -- & 312 & 1513 \\
        float8 & -- & 624 & 3026 \\
        \bottomrule
    \end{tabular} 
    \end{minipage}\hfill
    \begin{minipage}[t]{0.4\linewidth}
    \centering
    \begin{tabular}{l cc}
        \toprule
        & $E_\mathrm{add}$ [pJ] & $E_\mathrm{mul}$ [pJ] \\
        \midrule
        float32 & 0.38 & 1.31 \\
        bfloat16 & 0.11 & 0.21 \\
        int8 & 0.007&  0.07\\
        \bottomrule
    \end{tabular}
    \end{minipage}
    \end{small}
    \caption{Left: Floating-point operations per second (FLOPS) for NVIDIA Tensor Core GPUs on different data types. Right: Energy consumption per operation on 7nm process nodes, used in NVIDIA A100 GPUs~\cite{6757323,wang2023bitnet}.}
    \label{tab:flops}
\end{table}

The speed of GPU operations for all data types has been increasing continuously, as illustrated in Tab.~\ref{tab:flops}. From the rate of this acceleration we see that modern GPUs are increasingly optimized for smaller data types, such as bfloat16, float8 and int8 instead of the established single-precision float32. For scientific applications, most networks are still trained with float32 precision. If we want to reduce the computational cost of neural networks, we can first target the data type used throughout the high-dimensional latent space. This can significantly reduce computational cost, while maintaining performance, as demonstrated for bfloat16~\cite{kalamkar2019study}, int8~\cite{jacob2018quantization}, float8~\cite{micikevicius2022fp8}, and float4~\cite{liu2023llm}.

During training, we apply the automatic mixed precision strategy of using full precision for specific precision-sensitive operations, as well as the backward pass~\cite{micikevicius2017mixed}. We do not reduce the precision in input and output layers beyond bfloat16 to enable the network to embed and output high-precision numbers. However, we use int8 precision in the hidden linear layers, which dominate the computational cost. To further speed up training, we could also perform the backward pass at lower precision, but would have to use stochastic rounding~\cite{gupta2015deep}. In general, the push to lower-dimensional latent representations saturates once the network starts to map out the full range of the provided data type. In practice, this happens around int8 or float8 precision~\cite{wang2023bitnet,ma2024era}.

Our implementation emulates low-precision data types in linear layers, because linear layers operating on int8 are not widely supported yet. We use automatic mixed precision throughout the network and quantize the input of linear layers to the range of int8 values before performing the linear operation in bfloat16 precision. To quantize the inputs with minimal loss of information, we use zero-point quantization~\cite{jacob2018quantization}:
\begin{align}
    x_q &= \mathrm{clip}_{[q_{\min},q_{\max}]}\left(\left\lfloor \frac{x}{s} \right\rceil + z\right) \; ,
    \qquad
    x \;\approx\; s\,(x_q-z) \; .
    \label{eq:zeropoint_quant}
\end{align}
The scale $s$ maps the range $[x_\text{min},x_\text{max}]$ to $[q_\text{min},q_\text{min}]$ ($[-128, 127]$ for int8) and the zeropoint $z$ centers the range to align the zero values. $\lfloor\cdot\rceil$ denotes rounding and $\mathrm{clip}_{[q_{\min},q_{\max}]}$ applies the min and max functions with the quantization range. The values $s$ and $z$ can be computed on the fly for each tensor (dynamic quantization) or pre-computed from training data and fixed at inference (static quantization). We only use static quantization for the small taggers of Section~\ref{sec:fpga-tagging}, as the cost of dynamic quantization is negligible for large networks. We find similar performance when replacing int8 by float8, but prefer int8 because of the slightly reduced computational cost. Weights are quantized in the same manner. We use straight-through estimation to propagate gradients through this non-differentiable quantization operation. For our four architectures this means:
\begin{itemize}
    \item Transformer: We first use automatic mixed precision throughout. For all linear layers except the input and output projections, we quantize inputs to int8 using straight-through estimation.
    \item ParT: We again use automatic mixed precision on the whole architecture. Linear layers in the particle attention and class attention blocks are quantized to int8 in the same manner as above. The feed-forward networks embedding particle-wise and pair-wise features are also quantized to int8, except their first layer which uses bfloat16 precision.
    \item L-GATr-slim: We apply automatic mixed precision and int8 quantization just as for the transformer. 
    \item LLoCa-Transformer: The LLoCa-Transformer uses the vanilla transformer as its backbone, and we apply automatic mixed precision and int8 quantization accordingly. The linear layers in the Frames-Net are quantized to int8, whereas the orthonormalization procedure is performed in full float32 precision, to avoid numerical symmetry violation. The frame-to-frame transformations in attention are also performed in float32 precision.
\end{itemize}
%

\subsection{Weight quantization}
\label{sec:weight-quantization}

To reduce the computational cost further we can quantize the linear weights to binary or ternary values, while keeping activations in higher precision~\cite{wang2023bitnet,ma2024era}. These quantized weights are equipped with a full-precision scale factor that is shared across the full layer. Matrix multiplication with these binary or ternary weights reduces to addition, which can be implemented very efficiently. In addition, quantized weights can be stored more efficiently because of the reduced memory cost.

Quantization-aware training (QAT) is challenging, as gradient descent relies on continuous weight updates, whereas quantized weights inherently lead to non-differentiable losses. We start from a generic loss function and formulate weight quantization as an additional regularization constraint that yields $R(\theta_k)=0$ for the quantized parameters,
\begin{align}
  \hat\theta
  = \argmin_\theta \loss_\text{QAT}(\theta)
  \qquad\text{with}\qquad
  \loss_\text{QAT}(\theta )= \loss(\theta) + \lambda \sum_k R(\theta_k)\; .
    \label{eq:qat_objective}
\end{align}
Minimizing this combined loss is impractical. The proximal gradient method instead implements the effect of $R(\theta_k )$ as a modification to the gradient descent update rule~\cite{bai2018proxquant}, such that the updated parameters $\theta'$ minimize Eq.~\eqref{eq:qat_objective}. We expand the loss to second order
\begin{align}
  \loss(\theta' )
  \approx \loss(\theta )
  + \sum_k\frac{\partial\loss(\theta)}{\partial\theta_k}(\theta'_k - \theta_k)
  + \frac{1}{2}\sum_{k,l} (\theta'_k-\theta_k) \frac{\partial^2\loss(\theta)}{\partial\theta_k \partial\theta_l} (\theta'_l-\theta_l)\;.
\end{align}
With the standard Lipschitz assumption for the loss, we can bound the second derivative with a sufficiently large constant $\kappa$,
\begin{align}
  \sum_{k,l} (\theta'_k-\theta_k) &\frac{\partial^2\loss(\theta)}{\partial\theta_k \partial\theta_l}(\theta'_l-\theta_l) \le \kappa \left| \theta'-\theta \right|^2 \notag \\
  \Rightarrow \quad 
  \loss(\theta') &\le \loss(\theta ) + \sum_k\frac{\partial\loss(\theta)}{\partial\theta_k}(\theta'_k - \theta_k) + \frac{\kappa}{2} \left|\theta' - \theta\right|^2 \notag\\
  &= \loss(\theta )
  + \frac{\kappa}{2}\sum_k\left[ \theta'_k - \left(\theta_k - \frac{1}{\kappa} \frac{\partial\loss(\theta)}{\partial\theta_k}\right)\right]^2
  - \frac{1}{2\kappa} \sum_k \left(\frac{\partial\loss(\theta)}{\partial\theta_k}\right)^2\; .
\end{align}
This gives us for the training objective in Eq.~\eqref{eq:qat_objective}
\begin{align}
    \hat\theta_k &= \argmin_{\theta_k'} \loss_\text{QAT}(\theta_k') \notag\\
    &= \argmin_{\theta_k'}
    \left[ \frac{\kappa}{2}\left[ \theta_k' - \left(\theta_k - \frac{1}{\kappa}\frac{\partial\loss(\theta)}{\partial\theta_k}\right)\right]^2
    + \lambda R(\theta_k') \right] \notag\\ 
    &= \argmin_{\theta_k'} \left[
      \left[\theta_k' - \left(\theta_k - \frac{1}{\kappa}\frac{\partial\loss(\theta)}{\partial\theta_k}\right)\right]^2
      + \frac{2\lambda}{\kappa} R(\theta_k') \right] \; ,
\end{align}
because adding and multiplying constants do not change the result of $\argmin$. The first term describes the standard gradient descent for $\loss(\theta)$ with a learning rate $1/\kappa$, leading to the naive parameter update
\begin{align}
    \tilde\theta_k = \theta_k - \frac{1}{\kappa} \frac{\partial\loss(\theta)}{\partial\theta_k}  \; .
\end{align}
We now define the proximal operator as a post-hoc modification of this update rule that yields the optimally updated parameters $\hat\theta$ from Eq.~\eqref{eq:qat_objective},
\begin{align}
  \hat\theta_k
  &= \text{prox}_R(\tilde\theta_k) \notag \\
  &\equiv \argmin_{\theta_k'} \left[ (\theta'_k - \tilde\theta_k)^2 + \frac{2 \lambda}{\kappa} R(\theta'_k) \right] \;.
\label{eq:prox_operator}
\end{align}
Instead of directly using the result from naive gradient descent $\tilde\theta_k$ to update the network parameters as $\theta'_k = \tilde\theta_k$, the proximal operator performs a post-processing on $\tilde\theta_k$ to account for the regularization term, with $\lambda/\kappa$ being a measure of the regularization strength.

\begin{figure}[t]
    \includegraphics[width=0.32\linewidth]{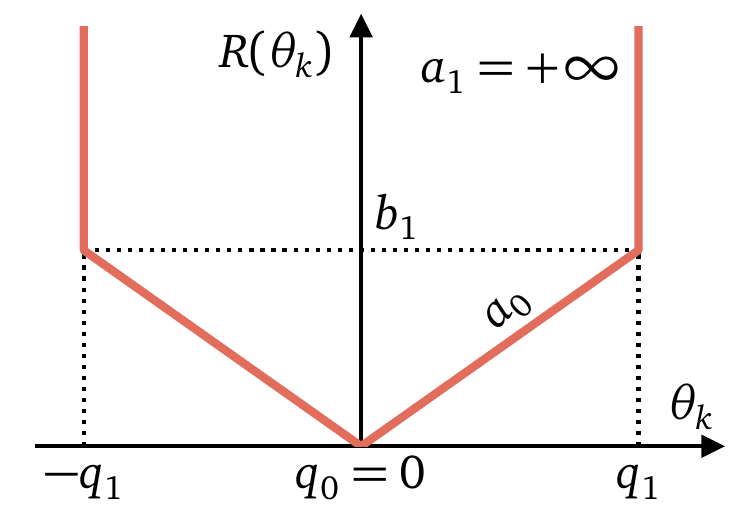}
    \includegraphics[width=0.32\linewidth]{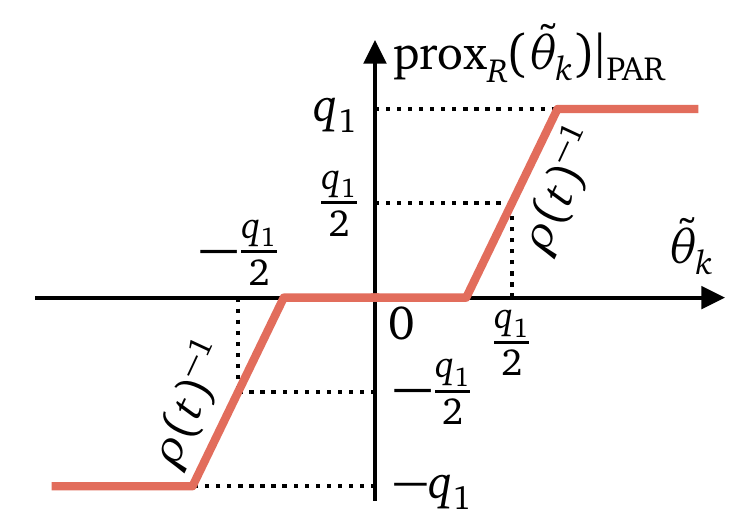}
    \includegraphics[width=0.32\linewidth]{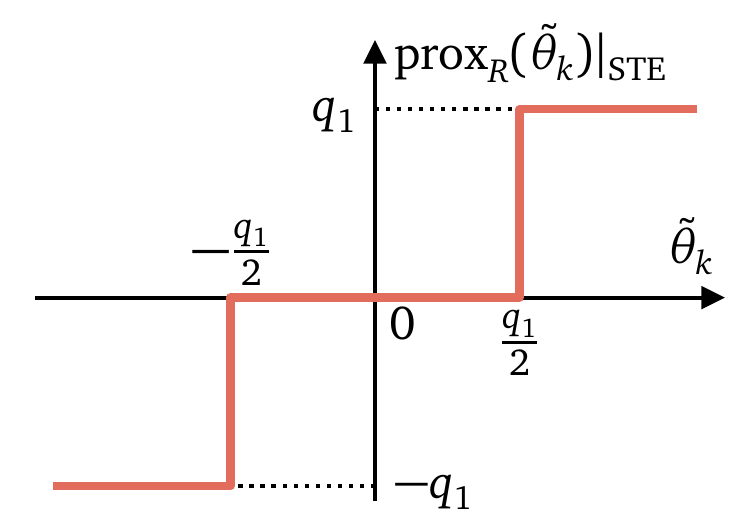}
    \caption{Regularizer function $R(\theta_k)$ for ternary quantization (left), proximal map $\text{prox}_R(\tilde{\theta}_k)$ for soft quantization at time $t$ (middle), and final proximal map for hard quantization (right). The proximal map parametrized by $\rho(t)$ corresponds to the regularizer function defined with $a_0 = (1+\rho(t)) q_1/2$ and $b_1 = a_0 q_1$.}
    \label{fig:illustrate_parq}
\end{figure}

An efficient implementation requires a closed form for the proximal operator. We write the regularization term as a piecewise-affine regularizer (PAR)~\cite{Jin:2025parq},
\begin{align}
    R(\theta_k) \Bigg|_\text{PAR} = \max_i \left[ a_i \left( |\theta_k | - q_i \right) + b_i\right] \; .
\end{align}
The slopes satisfy $0\le a_0<a_1<\cdots <a_m=\infty$, and for the quantized values $\{0, \pm q_1, \dots , \pm q_m\}$ the biases become
\begin{align}
  b_0=0
  \qquad \text{and} \qquad
    b_i=b_{i-1}+a_{i-1} \left( q_i-q_{i-1} \right) \; .
\end{align}
This form for $R(\theta_k)$ is visualized in Fig.~\ref{fig:illustrate_parq}. The proximal operator has a closed form
\begin{align}
    \text{prox}_R (\tilde\theta_k) \Bigg|_\text{PAR} = \begin{cases}
        \sign (\tilde\theta_k )q_i & |\tilde\theta_k| \in [a_{i-1}+q_i, a_i+q_i] \\[2mm]
        \tilde\theta_k - \sign (\tilde\theta_k)a_i & |\tilde\theta_k| \in [a_i+q_i,a_i+q_{i+1}]
    \end{cases}  \; .
\end{align}
We choose the parameters $\{a_i\}_{0\leq i \leq m}$ such that the slopes of the proximal map affine pieces are defined by a single parameter $\rho(t)$ depending on time or iteration count~\cite{Jin:2025parq}
\begin{align}
    a_i = \frac{1+\rho(t)}{2}q_{i+1} + \frac{1-\rho(t)}{2}q_i \; .
\end{align}
The parameter $\rho(t)$ corresponds to the inverse slope of the proximal map non-horizontal segments. During training, it is annealed from one to zero, with a sigmoid-like scheduler. This way the proximal map smoothly transitions from the identity to a step function, and the weights are progressively quantized. The evolution of the proximal map is illustrated in Fig.~\ref{fig:illustrate_parq}.

A straightforward quantization-aware training, straight-through estimation (STE), is a special case of the proximal gradient formalism where the regularizer consists of delta distributions, with a step-function as the proximal map. It is equivalent to replacing the annealing schedule with $\rho(t) = 0$.
\begin{align}
  R(\theta_k ) \Bigg|_\text{STE}
  &= \sum_i \left[ \delta (\theta_k - q_i) + \delta (\theta_k +q_i)\right] \notag \\
  \text{prox}_R (\tilde\theta_k ) \Bigg|_\text{STE}
  &= \sign (\tilde\theta_k) q_k\;.
\end{align}
We compare PARQ and STE for our applications and find that PARQ stabilizes training and achieves marginally better performance.

We use weight quantization for the same linear layers that use low-precision data types described in Sec.~\ref{sec:input-quantization}, with the exception of the LLoCa-Transformer Frames-Net where we keep weights at the same precision as the inputs. The fraction of quantized weights varies between 90\% and 99\% for the different architectures and numbers of parameters. 

\subsection{Quantized equivariant taggers}
\label{sec:gpu-tagging}

First, we compare quantization-aware training with PARQ and with the naive STE in Tab.~\ref{tab:ste-vs-parq}. The vanilla transformer maintains performance up to int8 precision, but drops in performance after weight quantization. ParT maintains performance all the way down to ternary weights. The two Lorentz-equivariant networks undergo slight performance drops when reducing precision to int8, and when quantizing weights to ternary values. Here, it is important to ensure that weight quantization and input quantization violate the Lorentz equivariance in a minimal way. For the LLoCa-Transformer, we perform the equivariance-critical orthonormalization and the projection into the local frames in full precision, and limit low-precision operations to Lorentz-invariants. For L-GATr-slim, we multiply full vectors with quantized weights, which again does not incur any additional Lorentz violation. Nevertheless, our quantized LLoCa-Transformer and L-GATr-slim can only satisfy Lorentz-equivariance to the precision of the linear layer inputs, which are float8, translating into violation effects at the 10\% level. 

\begin{table}[b!]
    \centering
    \begin{small}
    \begin{tabular}{l llll}
        \toprule
        & \multicolumn{4}{c}{Background rejection $1/\epsilon_B$ @ $\epsilon_S=0.3$} \\
        & f32 & i8 & i8+STE & i8+PARQ \\
        \midrule
        Transformer & \result{1694}{\dz{}69} & \result{1577}{\dz{}68} & \result{1357}{\dz{}69} & \result{1302}{\dz{}55}\\
        ParT & \result{1808}{\dz{}33} & \result{1768}{\dz{}81} & \result{1676}{\dz{}94} & \result{1693}{\dz{}86} \\
        LLoCa-Transformer & \result{2150}{130} & \result{2019}{\dz{}79} & \result{1932}{132} & \result{1935}{110}\\
        L-GATr-slim & \result{2264}{\dz{}93} & \result{2082}{124} & \result{1990}{108}& \result{1872}{\dz{}67}\\
        \bottomrule
    \end{tabular}
    \end{small}
    \caption{Impact of different kinds of input and weight quantization on network performance}
    \label{tab:ste-vs-parq}
\end{table}
\begin{figure}[b!]
    \centering
    \includegraphics[width=0.495\linewidth]{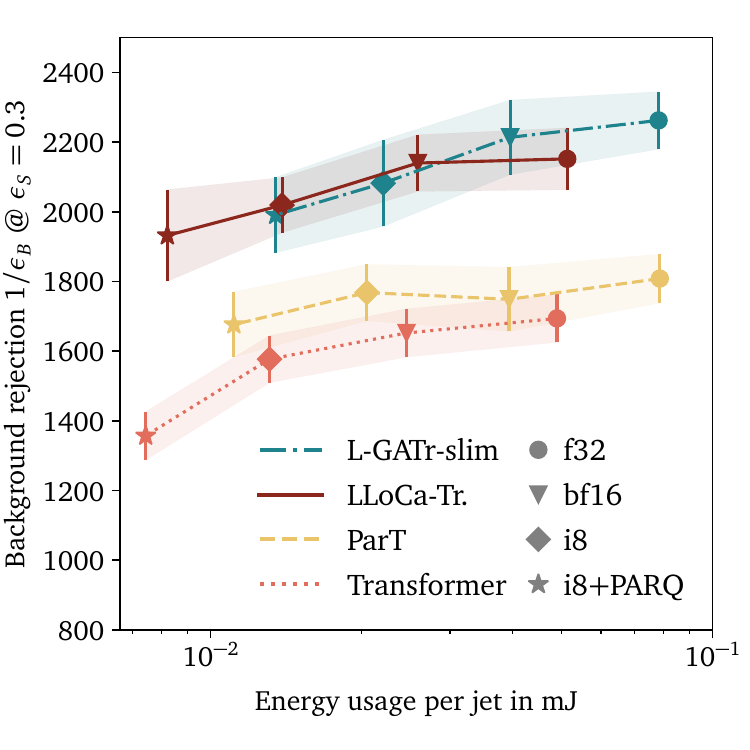}
    \caption{Energy consumption per jet using different taggers and quantization approaches.}
    \label{fig:energy-estimate}
\end{figure}

For top tagging with resource-efficient networks we estimate the energy use following Refs.~\cite{6757323,wang2023bitnet}. We use estimates for the energy cost per operation in Table~\ref{tab:flops}, and estimate the total energy consumption of a forward pass by counting all operations in the network and weighting them with the corresponding energy cost. We quantize inner linear layers to ternary weights, and reduce the precision of the latent representation as described in Sec.~\ref{sec:input-quantization}.

Next, we correlate the energy consumption with the performance of different quantized taggers in Fig.~\ref{fig:energy-estimate}. Our estimates for the energy cost indicate that every quantization step reduces the cost by a factor of about two, accumulating to an order-of-magnitude gain when combining all steps. L-GATr-slim consumes about 50\% more energy than a vanilla transformer due to its more expensive linear layers, whereas the LLoCa-Transformer incurs a negligible additional energy cost. Compared with the vanilla Transformer, ParT incurs about 50\% overhead with roughly equal contributions from preprocessing the pairwise features and from the larger MLP hidden dimension in the Transformer blocks. Importantly, these energy savings come at almost no performance loss, they only require the implementation of optimized low-precision operations.

\subsection{Towards online jet tagging}
\label{sec:fpga-tagging}

\begin{figure}[b!]
    \includegraphics[width=0.495\linewidth,page=1]{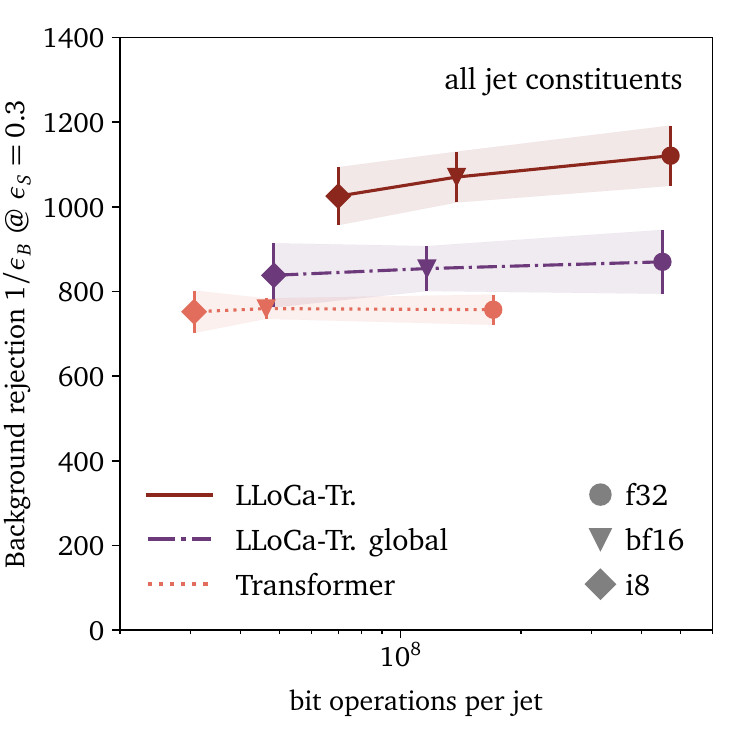}
    \includegraphics[width=0.495\linewidth,page=2]{figs/bitops.pdf}
    \caption{Number of bit-operations per jet for different taggers and quantization approaches, using the full range of jet constituents (left) and only up to 30 constituents (right).}
    \label{fig:bitops-estimate}
\end{figure}

An exciting question is whether small and quantized transformers can meet the resource and latency requirements of FPGAs. In that case, resource-efficient versions of the leading jet taggers can be applied at the trigger level. Information about the partonic nature of a jet would open entirely new avenues for HL-LHC analyses. The state-of-the-art approach to FPGA-ready networks is high granularity quantization (HGQ)~\cite{Sun:2024soe}, where we learn the bit-width of each operation using backpropagation. Since this study can only serve as a first proof of principle, we use GPU implementations instead of an FPGA, as well as our naive i8+PARQ quantization instead of HGQ.

For this test, we reduce a vanilla transformer and the LLoCa-Transformer to one block and a, typically, 16-dimensional latent space. Even for low numerical precision the orthonormalization has to be done to full precision, so its cost is significant. Therefore, we also show results for the global canonicalization LLoCa-Transformer, which requires the orthonormalization only for one particle, and also does not require frame-to-frame conversions in attention. From Fig.~\ref{fig:scaling} we know that L-GATr and L-GATr-slim show poor performance for a minimal number of blocks. Adding more blocks solves this problem, but significantly increases the latency, so we omit the L-GATr architecture for now. 

The results are shown in Fig.~\ref{fig:bitops-estimate}. In the left panel we see that we can reduce the numerical precision without significant performance drop, down to int8 with static quantization. We find that naively quantizing to ternary weights significantly degrades performance, and conclude that HGQ is necessary to improve further.
In the right panel of Fig.~\ref{fig:bitops-estimate} we reduce the input data from the full set of particle-flow constituents to the 30 constituents with largest transverse momentum. The number of constituents inside a top jet peaks around 50~\cite{Butter:2017cot}. This means that while the general picture does not change, the performance of the top taggers decreases significantly when we only consider the leading 30 constituents.

\section{Outlook}
\label{sec:conc}

As for industry applications, upscaled transformers are currently driving the performance frontier for jet tagging as the leading ML-application in LHC physics. However, this performance gain comes with an over-proportional compute cost in training and evaluation. For dedicated physics applications, we studied the opposite approach, physics-aware downscaling of transformers, in terms of reducing network parameters and using quantized parameters.

As a starting point, we have presented an efficient version of the equivariant L-GATr architecture, L-GATr-slim. It constructs powerful latent representations for many LHC tasks, from jet tagging to amplitude regression and event generation. Compared to L-GATr, we reduced the training time by a factor six and the memory footprint by a factor 2, at no cost in tagging performance. In this way, the specialized layer approach is competitive with the local canonicalization approach, specifically the LLoCa-Transformer.

For L-GATr-slim and the LLoCa-Transformer we then tested the performance down to 1k network parameters, either reducing the number of layers or reducing the number of parameters per layer. While this down-scaling comes with a reduced performance, the LLoCa-Transformer remains a powerful tool for a small number of blocks, while L-GATr-slim leads the competition for a fixed number of blocks and a reduced number of parameters per block.

To further improve efficiency, we quantized the network parameters using the standard STE and PARQ methods. The inductive bias from Lorentz equivariance prevents large performance drops from this weight quantization, such that the performance of the LLoCa-Transformer and L-GATr-slim are only mildly affected by input and by weight quantization. Here, it is crucial that we ensure that the weight quantization does not affect the Lorentz equivariance significantly. We have estimated that quantization can reduce the energy consumption by 10$\times$.

Finally, we have combined size-reduction and quantization strategies to construct a minimal Lorentz-equivariant tagger, with the ultimate goal of jet tagging at the trigger level. Even though the FPGA-inspired minimal LLoCa-Transformer loses a factor of two in background rejection compared to its full-size counterpart, it still outperforms all pre-graph top taggers~\cite{Kasieczka:2019dbj}.

\subsection*{Code availability}

Our code is available at \url{https://github.com/heidelberg-hepml/tagger-quantization}. The amplitude regression and event generation experiments were performed using \url{https://github.com/heidelberg-hepml/lloca-experiments}. L-GATr-slim is available as part of the L-GATr package \url{https://github.com/heidelberg-hepml/lgatr}.

\section*{Acknowledgements}

We would like to thank Thea Arrestad for expert advice on FPGAs and to Luigi Favaro, Víctor Bresó Pla, Huilin Qu and Ramon Winterhalder for many helpful discussions. This work is supported by the Deutsche Forschungsgemeinschaft (DFG, German Research Foundation) under grant
396021762 – TRR 257 \textsl{Particle Physics Phenomenology after the Higgs Discovery}. It has received funding from the European Union’s Horizon Europe research and innovation programme under the Marie Sklodowska-Curie grant agreement No 101168829, \textsl{Challenging AI with Challenges from Physics: How to solve fundamental problems in Physics by AI and vice versa} (AIPHY). JS gratefully acknowledges support from the Alexander-von-Humboldt foundation as
a Feodor-Lynen Fellow. The authors acknowledge support by the state of Baden-Württemberg through bwHPC and the German Research Foundation (DFG) through the grants INST 35/1597-1 FUGG and INST 39/1232-1 FUGG. This work made use of the facilities of the N8 Centre of Excellence in Computationally Intensive Research (N8 CIR) provided and funded by the N8 research partnership and EPSRC (Grant No. EP/T022167/1). The Centre is coordinated by the Universities of Durham, Manchester and York.

\appendix
\section{Implementation details}

\label{app:hyperparams}
We use the github repositories \url{https://github.com/heidelberg-hepml/tagger-quantization} for jet tagging, and \url{https://github.com/heidelberg-hepml/lloca-experiments} for amplitude regression and event generation. All hyperparameter choices can be found in the \texttt{config/} folders which is present in both repositories, and the file \texttt{REPRODUCE.md} contains the commands to reproduce all results given in this paper.

\subsection*{Jet tagging}

Our LLoCa-Transformer closely follows Ref.~\cite{Favaro:2025pgz}, with a smaller Frames-Net that uses 32 instead of 128 hidden channels in the MLP. We find that this setting achieves the same performance, but reduces FLOPs and memory use significantly. Apart from this change, our vanilla transformer and LLoCa-Transformer trainings use the same architecture and training hyperparameters as Ref.~\cite{Favaro:2025pgz}.

Our L-GATr-slim tagger uses 12 blocks, 32 vector channels, 96 scalar channels, and 8 attention heads. It is trained for 200k iterations on the top tagging dataset, using the Lion optimizer with batch size 128, learning rate $3\times 10^{-5}$, weight decay 2, and a cosine annealing learning rate schedule. The JetClass training uses 1M iterations, and using the same optimizer, batch size, learning rate, and learning rate schedule as ParT, see Ref.~\cite{Qu:2022mxj}. To finetune a L-GATr-slim tagger that was pretrained on the JetClass dataset using the same training setup, we use the training setup described above with a learning rate of $3\times 10^{-6}$ for the transformer backbone, and a learning rate of $3\times 10^{-4}$ for the final linear layer.

For our ParT trainings, we use the same improved training hyperparameters as for L-GATr-slim, instead of the original training setup~\cite{Qu:2022mxj}. We find that these hyperparameters achieve significantly better results, see Tab.~\ref{tab:top_tagging}.

\subsection*{Amplitude regression}

Our L-GATr-slim amplitude regressor is designed to closely follow the shape of the L-GATr amplitude regressor in Ref.~\cite{Favaro:2025pgz}, with 8 blocks, 40 vector channels, 72 scalar channels, and 8 heads. It follows the training setup used for all amplitude networks in Ref.~\cite{Favaro:2025pgz}, with 200k iterations, Adam optimizer, batch size 1024, learning rate $10^{-3}$, and a step-wise learning rate reduction whenever the loss saturates.

\subsection*{Event generation}

The L-GATr-slim event generator closely follows the L-GATr event generator in Ref.~\cite{Brehmer:2024yqw}, with 6 blocks, 32 vector channels, 64 scalar channels, and 8 heads. It uses the same training setup with 200k iterations, AdamW optimizer, batch size 2048, and learning rate $10^{-3}$, and a step-wise learning rate reduction whenever the loss saturates.

\subsection*{Ultra-mini taggers}

For each network in Fig.~\ref{fig:scaling}, we provide the hyperparameters of the architecture in Table~\ref{tab:mini-taggers}, with 'deep' networks corresponding to the right panel. We use different training hyperparameters for each network size, given in Table~\ref{tab:mini-training}. All networks are trained for 200 000 iterations using the Adam optimizer with a cosine-annealing learning rate schedule. We use the same training hyperparameters for the deep and shallow networks.

\subsection*{Quantized taggers}

The quantized networks use the same hyperparameters as their non-quantized counterpart. We replace the default two-cycle cosine annealing learning rate schedule for the Transformer and LLoCa-Transformer with a one-cycle cosine annealing schedule when using PARQ. This avoids increasing the learning rate when starting the quantization with the PARQ scheduler. The PARQ scheduler steepness~\cite{Jin:2025parq} is set to 100.

For the large taggers, int8 quantization is dynamic for both inputs and weights.
For the small taggers, we quantize the weights dynamically during training and we use static quantization for the inputs. The training starts with dynamic quantization, then the quantization parameters are saved after 10 000 steps and updated with each forward pass. To compute $s$ and $z$, we use the quantiles $10^{-3}$ and $1-10^{-3}$ of each input tensor as $x_\text{min}$ and $x_\text{max}$, and update the previous values with momentum $\beta=0.999$.

\begin{table}[tb]
    \centering
    \begin{tabular}{cc c ccc ccc}
    \toprule
    Architecture && 2M & 200k & 20k & 2k & 200k (deep) & 20k (deep) & 2k (deep)\\
    \midrule
    \multirow{3}{*}{Transformer} & channels & 128/256 & 64/128 & 32/64 & 16/32 & 32/64& 16/16& 4/4\\
    & heads & 8 & 4 & 4 & 2 & 4& 2&1  \\
    & blocks & 12 & 4 & 2 & 1 & 10& 10&10  \\
    \midrule
    FramesNet & channels & 32 & 16 & 8 & 4 & 16 & 8 & 4 \\
    (LLoCa)& layers & 2 & 2 & 2 & 2 & 2 & 2 & 2 \\
    \midrule
    \multirow{4}{*}{ParT} & channels & 128/512 & 64/128 & 32/64 & -- & 32/128& 8/16& 4/4\\
    & pair chan. & 64 & 16 & 8 & -- & 16 & 4 & 2 \\
    & heads & 8 & 4 & 2 & -- & 4& 1& 1\\
    & blocks & 8+2 & 3+1 & 1+1 & --  & 8+2& 8+2& 8+2\\
    \midrule
    \multirow{4}{*}{L-GATr-slim} & vectors & 32/128 & 16/32& 8/16& 4/4 & 8/32 & 4/4 & 2/2\\
    & scalars & 96/384& 64/128& 32/64& 16/16 & 32/128& 16/16& 4/4\\
    & heads & 8& 4& 4& 2 & 4& 2& 1\\
    & blocks & 12& 4& 2& 1 & 10& 10& 10\\
    \bottomrule
    \end{tabular}
    \caption{Architecture hyperparameters for the different taggers of Fig.~\ref{fig:scaling}. For ParT, we write N+M for N self-attention and M class-attention blocks. For channels, we write N/M for a N-dimensional latent space in attention, and a M-dimensional increased latent space used in the MLP for activations.}
    \label{tab:mini-taggers}
\end{table}

\begin{table}[tb]
    \centering
    \begin{tabular}{c ccc}
    \toprule
    & 200k & 20k & 2k\\
    \midrule
    Batch size & 256 & 512 & 4096 \\
    Learning rate & $10^{-3}$ & $3 \times 10^{-3}$ & $10^{-2}$ \\
    Weight decay & 0.1 & 0 & 0 \\
    \bottomrule
    \end{tabular}
    \caption{Training hyperparameters for the different taggers of Fig.~\ref{fig:scaling}. The same hyperparameters are used for the deep and shallow networks. The standard 2M networks use the same training configurations as their original publication.}
    \label{tab:mini-training}
\end{table}

\bibliography{tilman,literature}
\end{document}